\documentclass{iopart}
\usepackage{iopams}

\usepackage{longtable}
\usepackage{tabularx}

\usepackage[final]{graphicx}% Include figure files
\usepackage{dcolumn}% Align table columns on decimal point
\usepackage{bm}% bold math

\def\extgra{eps}

\newcommand{\mm}[1]     {\ifmmode {#1} \else{}${#1}$\fi}
\newcommand{\mmm}[1]    {\ifmmode{}#1 \else{}${#1}$\fi}
\newcommand{\beq}[1]    {\begin{equation} \label{#1}}
\newcommand{ \eeq}{\end{equation}}

\def\xfive{\mm{\rm{ \sqrt{5}\times\sqrt{5} } } }
\def\xtwo{\mm{\rm{ \sqrt{2}\times\sqrt{2} } } }

\def\fesex{\mm{\rm{ Fe_{2-x}Se_2Cs_{y} } } }
\def\fesex{\mm{\rm{ Cs_{y}Fe_{2-x}Se_2 } } }
\def\fesexx{\mm{\rm{ X_{y}Fe_{2-x}Se_2 } } }

\def\fesek{\mm{\rm{ K_{y}Fe_{2-x}Se_2 } } }
\def\feserb{\mm{\rm{ Rb_{y}Fe_{2-x}Se_2 } } }

\def\rbh{K43}
\def\rbl{K72}

\def\fese{\mm{\rm{ FeSe } } }

\def\Tc\mm{T_c}

\begin{document}

% \draft command makes pacs numbers print

\title[Intrinsic crystal phase separation in $\rm{Rb_{y}Fe_{2-x}Se_2}$]{Intrinsic crystal phase separation in antiferromagnetic superconductor $\rm\mathbf{Rb_{y}Fe_{2-x}Se_2}$: a diffraction study}

% repeat the \author\address pair as needed

\author{V.~Yu.~Pomjakushin$^{1}$,  A.~Krzton-Maziopa$^{2}$\footnote[3]{
On leave from: Faculty of Chemistry, Warsaw University of Technology, 00-664 Warsaw, Poland.}, E.~V.~Pomjakushina$^{2}$, K.~Conder$^{2}$, D.~Chernyshov$^{3}$, V.~Svitlyk$^{3}$, A.~Bosak$^{4}$ }
\address{$^{1}$ Laboratory for Neutron Scattering, Paul
Scherrer Institut, CH-5232
Villigen PSI, Switzerland}

\address{$^{2}$ Laboratory for Developments and Methods, PSI, CH-5232
Villigen PSI, Switzerland}

\address{$^{3}$ Swiss-Norwegian Beam Lines at ESRF, BP220, 38043
Grenoble, France}

\address{$^{4}$ European Synchrotron Radiation Facility, BP 220, 38043 Grenoble Cedex, France}

\ead{Vladimir.Pomjakushin@psi.ch}

%\pagebreak

\date{\today}

\begin{abstract}

The crystal and magnetic structures of the superconducting iron based chalcogenides \feserb\ have been studied by means of single crystal synchrotron x-ray and high resolution neutron powder diffraction in the temperature range 2-570~K. The ground state of the crystal is an intrinsically phase separated state with two distinct by symmetry phases. The main phase has the iron vacancy ordered \xfive\ superstructure ($I4/m$ space group) with AFM ordered Fe spins. The minority phase does not have \xfive-type of ordering and has smaller in plane lattice constant $a$ and larger tetragonal $c$-axis and  can be well described assuming the parent average vacancy disordered structure ($I4/mmm$ space group) with the refined stoichiometry  $\rm {Rb_{0.60(5)}(Fe_{1.10(5))}Se})_2$. The minority phase amounts to 8-10\% mass fraction. The unit cell volume of the minority phase is 3.2\% smaller than the one of the main phase at $T=2$~K and has quite different temperature dependence. The minority phase transforms to the main vacancy ordered phase on heating above the phase separation temperature $T_P=475$~K. The spatial dimensions of the phase domains strongly increase above $T_P$ from 1000~\AA\ to $>2500$~\AA\ due to the merging of the regions of the main phase that were separated by the second phase at low temperatures. Additional annealing of the crystals at the temperature $T=488$~K close to $T_P$ for the long time drastically reduces the amount of the minority phase.

\end{abstract}

% insert suggested PACS numbers in braces on next line

\maketitle

\section{Introduction}
The discovery of the Fe-based pnictide superconductors has triggered a remarkable renewed interest for possible new routes leading to high-temperature superconductivity. As observed in the cuprates, the iron-based superconductors exhibit interplay between magnetism and superconductivity suggesting the possible occurrence of unconventional superconducting states. Among the iron-based superconductors \fese\ has the simplest structure with layers in which Fe cations are tetrahedrally coordinated by Se \cite{Hsu2008}. Recently superconductivity at about 30K was found in chalcogenides \fesexx\  for X=K, Cs, Rb \cite{PhysRevB.82.180520,Krzton2010,2010arXiv10125525W}.  An average crystal structure of \fesexx\ is the same as in the layered (122-type) iron pnictides with the space group $I4/mmm$\cite{PhysRevLett.101.107006}. The principal difference of the new chalcogenides \fesexx\ (X=K, Tl, Rb, Cs)  is the presence of a superstructure due to the iron vacancy ordering and strong antiferromagnetism (AFM) with large iron magnetic moments 
\cite{Fang2010, pom2011fesex, pom2011feserb,Zavalij2011, Svitlyk2011, ZWang, Basca2011, Bao2011, Ricci,Ricci2, MWang, Cai2011, Kazakov,Shermadini2012,Weili2012,Friemel2012}.
 There is a general agreement on the presence of the $\sqrt{5}\times\sqrt{5}$ vacancy ordered structure and a second minority phase possessing different structure with additional reflections with propagation vector [$1\over2$,$1\over2$] that is often refereed as a $\sqrt{2}\times\sqrt{2}$ structure. 
An important question whether the antiferromagnetically and vacancy ordered state (AFMV) microscopically coexists with the superconductivity (SC) remains open. 
In the transmission electron microscopy experiment \cite{ZWang,Yuan2012}, the superstructure \xfive\ was observed together with \xtwo\ structure in certain areas in non-SC crystals of \fesek\ ($x=0.4-0.5$), whereas the SC crystals ($x=0.3-0.4$) showed a phase separated state along c-axis with \xfive\ superstructure and disordered 122-structure. In the focusing synchrotron diffraction experiments \cite{Ricci, Ricci2} the presence of phase separated competing phases is reported in SC \fesek: a majority \xfive phase and a minority phase (30\%) having an in-plane compressed lattice volume and \xtwo\ weak superstructure. The minority phase disappears on heating above 520~K. The compressed phase with \xtwo-superstructure has been also observed in the single crystal neutron diffraction experiments \cite{MWang}. This phase was tentatively described in $Pnma$ space group with an Fe vacancy model \cite{MWang}. Scanning tunneling microscopy studies of the local structural and electronic properties \cite{Cai2011} also show charge density modulation with \xtwo\ periodicity, but no famous Fe vacancies in SC \fesek, thus proposing a microscopic coexistence of SC state and an AFM state in \fesek\ without Fe-vacancies. 
Another scanning tunneling microscope measurement of the atomic and electronic structures of \fesek\ demonstrates that the sample contains two distinct phases: an insulating phase with well-defined  order of Fe vacancies, and a superconducting $\rm K(FeSe)_2$ phase containing no Fe vacancies\cite{Weili2012}.

In our previous work \cite{Bosak2012} we have also reported on the observation of the second minority in-plane compressed phase (MCP) phase with [$1\over2$,$1\over2$,L] Bragg rods in \fesex. In the present paper we present explicit evidences of intrinsic phase separation and the temperature evolution of the crystal structures of both phases in full details from T=1.5~K to 570~K in superconducting crystals of \feserb. 

\section{Samples. Experimental}
\label{exp}
Single crystals of rubidium intercalated iron selenides of nominal compositions $\rm {Rb_{0.85}(FeSe_{0.98}})_2$ and $\rm {Rb_{0.85}(Fe_{0.9}Se})_2$, denoted as \rbh\ and \rbl\,  were grown from the melt using the Bridgman method as described in Ref.~\cite{Krzton2010}.  Differential scanning calorimetry (DSC) experiments were performed with a Netzsch DSC 204F1 system. Measurements were performed on heating and cooling with a rate of 10 K/min using 20 mg samples encapsulated in standard Al crucibles. An argon stream was used during the whole experiment as protecting gas. Neutron powder diffraction (NPD) experiments were carried out at the SINQ spallation source of Paul Scherrer Institute (Switzerland) using the high-resolution diffractometer for thermal neutrons HRPT \cite{hrpt}. The samples used in the NPD experiments at low (below room temperatures using a cryostat) and high (with a furnace) temperatures were \rbl\ and \rbh, respectively. The samples for the NPD were prepared by pulverization of relatively large (100-200~mg) pieces of the single crystals in an inert atmosphere. The superconducting state has been identified by ac-susceptibility measurements with small pieces of the crystals using a conventional PPMS magnetometer. The onset of the diamagnetic response amounted to $T_{c}$ = 27 K and 24 K for \rbl\ and \rbh\ samples, correspondingly. We note that for the diffraction measurements significantly bigger amounts of crystals were used in comparison with the macroscopic measurements.
The low temperature macroscopic magnetic properties we studied in details in Ref. \cite{Boska2012} by means of superconducting quantum interference device (SQUID) and torque magnetometry with the crystal identical to the crystal \rbl\ of the present paper. Both crystals had the same $T_{c}$ and chemical composition determined by micro-XRF technique \cite{Anka2012}. 
Additionally we have prepared two more samples from the original \rbl\ and \rbh\ crystals. The fresh pieces of both crystals \rbl\ and \rbh\ were annealed in closed ampoules during 100~h and 50~h, respectively at temperature $T=488$~K in the vicinity of the phase separation temperature $T_P$. The annealed versions of samples were studied by NPD only at room temperature (section \ref{sec_an}).

The refinements of crystal and magnetic structures from neutron powder diffraction data were carried out with {\tt FULLPROF}~\cite{Fullprof} program, with the use of its internal tables for scattering lengths and magnetic form factors. Single crystal diffraction data were collected  at the SNBL beamline BM1A at the ESRF synchrotron in Grenoble (France) with a MAR345 image-plate area detector using $\lambda= 0.6977(1)$~\AA. Intensities were indexed and integrated with {\tt CrysAlis}\cite{crysalis}, empirical absorption correction was made with SADABS \cite{sadabs}, structure refinement with {\tt SHELXL97}\cite{shel}.

\begin{figure}[t]
  \begin{center}
    \includegraphics[width=8.5cm]{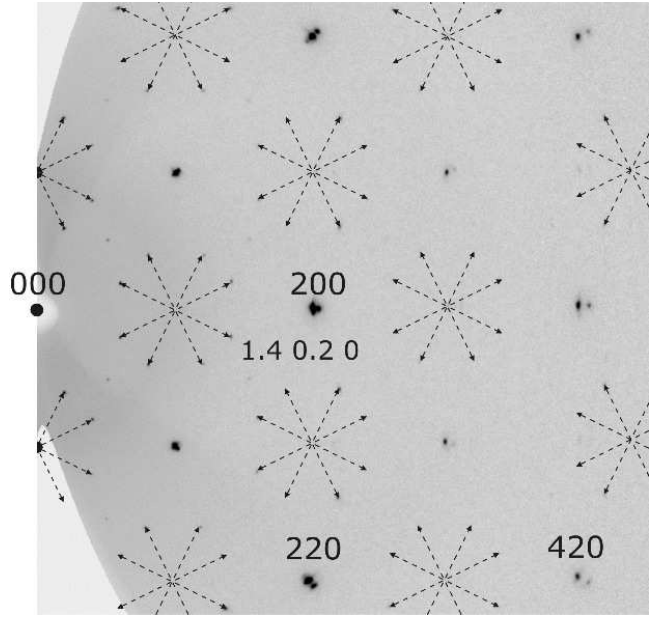} %{figure1_hk0.\extgra} 
  \end{center}

\caption{A slice of the reciprocal space ($\pm0.1c^*$) showing [hk0] plane at room temperature from single crystal x-ray diffraction experiment. The indexing is given in the average cell ($I 4/mmm$). The superstructure satellite reflections are indicated by arrows for propagation vectors {\bf k}$_1=[{2\over5},{1\over5},1]$ and  {\bf k}$_2=[{1\over5},{2\over5},1]$. These two k vector stars correspond to two twin domains  \cite{pom2011fesex}.}
\label{hk-plane}
\end{figure}

\section{Results and discussion}
\label{res}

\subsection{Phase separation and symmetry of the phases}

Figure \ref{hk-plane} shows a slice of the reciprocal space near $[hk0]$ plane. The superstructure reflections belonging to two twin domains can be easily identified. Each twin domain is described by a four arms star as discussed in Ref.~\cite{pom2011fesex}. The propagation vector star spanned by the propagation vector $k_1=[{2\over5},{1\over5},1]$ corresponds to the new 5 times bigger unit cell given by the basis transformation {\bf A}=2{\bf a}+{\bf b},  {\bf B}= -{\bf a}+2{\bf b},  {\bf C}= {\bf c}, where the lower case letters stand for the basis of the parent average $I4/mmm$ structure with $a\sim4, c\sim15$\AA. The capital letters denote the basis of the vacancy ordered AFMV phase with $I4/m$ structure \cite{pom2011fesex}. There is some confusion in the literature regarding the k-vector value. We note that in the centered lattices the propagation vector can contain integer components because the centered basis is larger than primitive. In the present case of $I$-centered latice adding an integer $1$ to only one of its components does not transform the propagation vector to an equivalent one, e.g. $[{2\over5},{1\over5},0]$ is not equivalent to $k_1$. One can choose an equivalent to $k_1$ propagation vector $k'_1 =[-3/5, 1/5,0]$. In the primitive basis both $k_1$ and $k'_1$ correspond to the same unique propagation vector $k_{1p}= [2/5,-2/5,4/5]$, which is $k2=[\mu,-\mu,\nu]$ in Kovalev notation~\cite{kovalev} or C-point $[\nu-\mu,\nu+\mu, 0]$ of primitive tetragonal BZ in CDML~\cite{Isotropy}.  

The single crystal diffraction shows (Fig. \ref{hk-plane}) that the parent average structure Bragg peaks are split similar as reported in \cite{Bosak2012}. The splitting is especially well seen at higher $hk$ indexes due to both better resolution and larger distance between the peaks, e.g. the peak $(420)$ in Fig.~\ref{hk-plane}. This splitting can be attributed to the presence of the minority in-plane compressed phase (MCP). The symmetry of the MCP was reported to be not higher than monoclinic due to the 4 fold splitting of (00L) peaks with large $L$ in $(a^*b^*)$-plane \cite{Bosak2012}.%  with an estimation of the distortion $\gamma\simeq90.6^\circ$

\begin{figure}[t]
  \begin{center}
    \includegraphics[width=8.5cm]{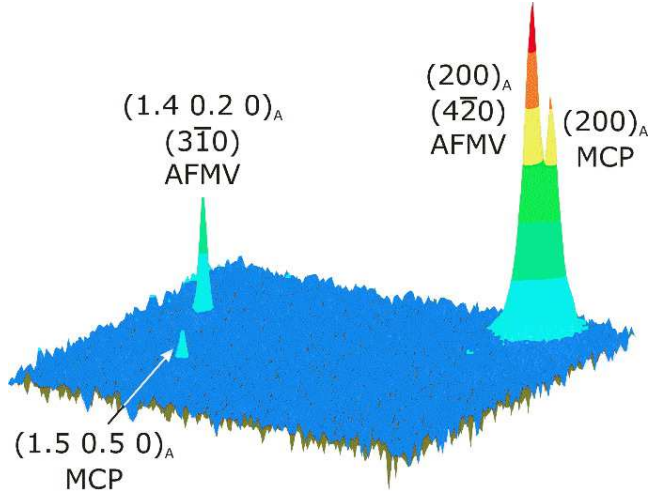}%{figure2_split.\extgra}
  \end{center}

\caption{Surface plot of the intensity distribution around parent Bragg peak (2,0,0)$_A$ (right) and the satellite at $(-1.4,0.2,0)$$_A$ (left). The z-axis has a logarithmic scale. One can see that the superstructure satellite peak shape is symmetric, whereas the parent peak is split. The intensities were obtained with integration along $c^*$ within the layer $\pm0.1c^*$.}
  \label{sat-split}
\end{figure}

Opposite to the main AFMV phase the MCP phase does not possess similar additional superstructure reflections. A careful inspection of the satellite peaks does not reveal any splitting. As an example, Fig. \ref{sat-split} shows a surface plot of the main Bragg peak (220)$_A$ [(4,-2,0)] and the satellite (1.4,0.2,0,0)$_A$ [(3,-1,0)] that are also indicated in Fig. \ref{hk-plane}. We use the underscore $A$ to denote the indexing in the average structure model $I4/mmm$, whereas the $hkl$-indexes without subscript refer to the 5 times bigger $I4/m$ unit cell. One can clearly see a splitting of the (200)$_A$ peak, but the shape of the satellite peak is symmetric. The crystals of \fesexx\ are quite fragile and often contain blocks. However the fact that the MCP has different symmetry allows one to conclude that the MCP is not simply a block of the main phase but a distinctly different crystal phase without 3D long range superstructure. There are superstructural Bragg rods along $c^*$ [the ($3\over2$,$1\over2$,0) section of the rod with integration $\pm0.1c^*$ is shown in Fig.~\ref{sat-split}] with in-plane propagation vector [$1\over2$,$1\over2$] that are commensurate with the minority compressed phase as shown in Ref.~\cite{Bosak2012}. These Bragg rods must originate from a 2D superstructure in tetragonal $(ab)$ plane in the MCP phase. The contribution of the Bragg rods to the powder diffraction patterns is too small to be taken into account. 

\begin{figure}[t]
  \begin{center}
    \includegraphics[width=9.3cm]{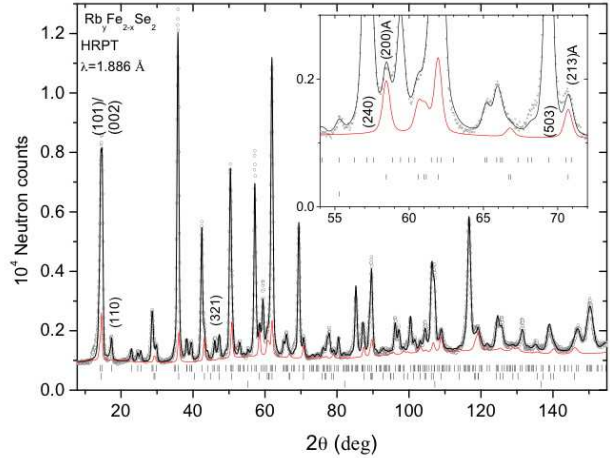} % fesex_riet
  \end{center}

\caption{The neutron diffraction pattern of \feserb\ (\rbl) at room temperature. Contribution of the minority in-plane compressed phase refined in the average crystal structure ({$ I4/mmm$} with $a=3.85, c=14.79$~\AA)  is shown by red solid curve. The upper row of tics indicates the Bragg peak positions from the structure of the main vacancy ordered phase, the middle hashmarks indicate reflections associated with the second phase, the bottom tics is Fe-impurity. The inset shows the explicit splitting of the parent Bragg peaks due to the presence of the second MCP phase. The $(200)_A$, $(213)_A$ Bragg peaks ($I4/mmm$) correspond to the $(240)$, $(503)$  peaks in the $I4/m$ structure, respectively.}
  \label{rietv-ND}
\end{figure}

Figure \ref{rietv-ND} shows the experimental NPD pattern and the Rietveld refined profile. There are two main phases in the refinement: the main AFMV phase with $I4/m$ crystal structure and $\tau_2$ ($I4/m'$) AFM order \cite{pom2011feserb} and the second MCP phase. There is little fraction of elemental Fe impurity ($<0.3$\% mass fraction), which can actually be disregarded in the refinements. Both the peak positions and intensities of the MCP phase are well described by the average $I4/mmm$ structure with smaller $a$ and larger $c$ lattice constants. In the powder diffraction pattern the MCP contribution is mainly overlapped with the peaks of the main phase, but there are some peaks that are very well separated due to the high resolution of NPD data. For instance (200)$_A$ [$(240)$] peak, which is also shown in Fig.~\ref{hk-plane}, shows a clear splitting and the peak (200)$_A$ from MCP is shifted to higher $2\theta$ because of the smaller $a$-constant as shown in the inset of Fig. \ref{rietv-ND}. 

The peaks of the main phase that overlap with the second phase peaks are those peaks of $I4/m$ structure that originate from the parent average phase. Disregarding the second phase in the analysis leads to an effective increase of the weight of the parent reflections and as a result underestimation of the superstructure reflections. In the single crystal diffraction experiment it is especially difficult to extract the true integrated intensity of the main phase for the Bragg reflections with small $hk$-indexes due to low resolution.

The true symmetry of the MCP phase might be lower than tetragonal \cite{Bosak2012}, but this deviation seems to be small to be accounted in the powder ND experiment for the minority phase with about 10\% contribution to the diffraction intensities. Anyway, we even do not have a model for the basis transformation that would account for the additional splitting of $(00L)$ peaks of the MCP phase in the (ab)-plane observed in the single crystal diffraction experiment \cite{Bosak2012}.

\begin{figure}[t]
  \begin{center}
    \includegraphics[width=9cm]{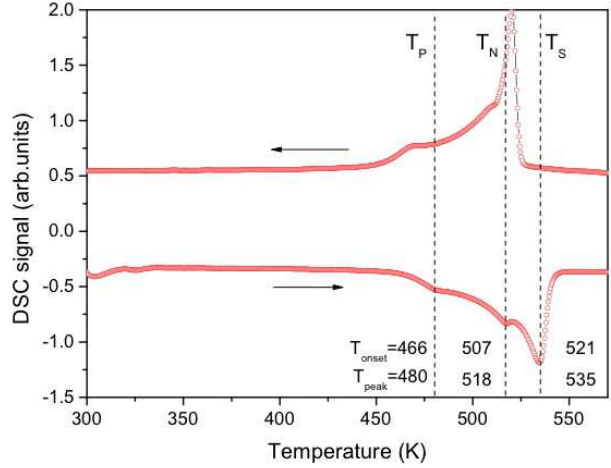} %fesex_voft_dsc
  \end{center}

 \caption{
Differential scanning calorimetry (DSC) signal as a function of temperature. Three peaks are observed: the largest at $T_s=535$~K corresponds to the structure phase transition due to the vacancy ordering, $T_N$ and $T_P$ are related to AFM ordering and phase separation, respectively.
}
  \label{dsc}
\end{figure}

\subsection{DSC and temperature dependencies of Bragg peaks}

The inspection of differential scanning calorimetry (DCS) signal reveals three peaks (Fig. \ref{dsc}). The peak at the highest temperature $T_S$ is associated with the structure transition to the vacancy disordered phase on heating. DSC is especially sensitive to the first order phase transitions due to the release of latent enthalpy. The second order phase transitions, like AFM ordering, can also be seen in DSC curves as smaller peaks due to the abrupt changes in heat capacity an the transition temperature. The middle peak at $T_N$ is associated with the AFM transition and the peak at lowest temperature $T_P$ is related to the phase separation transition as we show below.

\begin{figure}[t]
  \begin{center}
    \includegraphics[width=8cm]{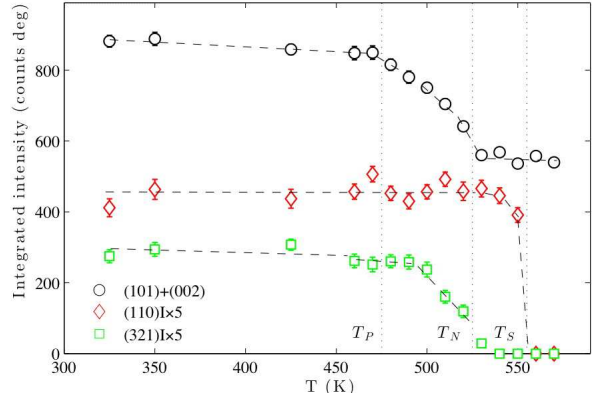} 
  \end{center}

 \caption{
Temperature dependences of selected neutron diffraction peaks. The vertical lines at $T_P$, $T_N$ and $T_S$ indicate phase separation, N\'{e}el temperatures and crystal structure transitions, respectively.}
  \label{NDpeaks}
\end{figure}

Figure \ref{NDpeaks} shows the temperature dependence of the integrated intensities of three selected neutron diffraction peaks indicated in Fig.~\ref{rietv-ND}. In the used magnetic model ($\tau_2$ or $I4/m'$ with the spins along the $c$-axis) the peak $(110)$ has purely crystal structure contribution. Note, that allowing spin components in $(ab)$ in the same $\tau_2$ symmetry would result in the appearance of the magnetic contribution in all superstructure peaks in general. The doublet $(101)/(002)$ has both magnetic and structure contribution. One can see that $(110)$ peak becomes abruptly zero above $T_s=555$~K and the doublet peak intensity flattens above N\'{e}el temperature $T_N=525$~K. The peak $(321)$ seems to have only magnetic contribution and vanishes above $T_N$. The transition temperatures seen by NPD might be different from the ones seen by DSC due to the temperature gradient between the thermocouple, which are mounted on the outer side of the vanadium container and the sample. Alternatively the difference might be due to different doping level, as noted in Ref.~\cite{Liu2011} where higher values of $T_N$ and $T_C$ were observed. Not exactly the same piece of sample was used for DSC and NPD. 

\subsection{Structure model}

The structure of the main phase is the vacancy ordered $I4/m$, whereas the structure of the second MCP is vacancy disordered $I4/mmm$ with the structure parameters listed in table \ref{tab1}. The fraction of the second phase amounted to 10\% and  8\% for the samples \rbl\ and \rbh, respectively. Since the minority phase fraction is small we used a constrained structure model.  The atomic displacement parameters (ADP) were chosen to be the same in both phases for the same types of elements. In the refinement of the data collected above room temperature the occupancies of the Rb and Fe of the MCP were fixed. The occupancy of the Rb2 site (2a) is refined to the larger than one values for the lower statistics ND patterns that were collected during the temperature scans. The contribution of the Rb2 site to the structure factor is four times smaller due to small Rb2 site multiplicity. The refinements in the model assuming equal occupancies of both Rb-sites (i.e. ideal disorder over the Rb sites) give only slightly worse reliability factors and similar structure parameters of both phases. The diffraction peak line shape parameters that are responsible for the size and microstrain broadening effects were constrained to be the same for both phases. This is a fair approximation keeping in mind that the second phase amounts to less that 10\% and disappears on heating. One can see from Fig.~ \ref{rietv-ND} that the peak shape of the MCP phase is well described under the above assumption. The magnetic model is the block spin antiferromagnetic  structure $\tau_2$ ($I4/m'$) with the spins on Fe2 site aligned along $c$-axis \cite{pom2011feserb}. The presence of the magnetic moment on Fe1 is forbidden by $\tau_2$ symmetry. The contribution of the Fe1 site to the diffraction intensity as well as to the total Fe stoichiometry is much smaller than the contribution of the Fe2 sites due to the four times smaller number of Fe1 atoms. However, the refined site occupancy of Fe1 is not very small and amounts to about 16\%. We have also observed substantial occupancy of Fe1 (4d) site in our previos single crystal diffraction experiment \cite{pom2011fesex}.  

The secondary MCP phase has smaller refined Rb-stoichiometry than in the main phase and no vacancies on Fe-site with the formula $\rm {Rb_{0.60(5)}(Fe_{1.10(5))}Se})_2$ (Table~\ref{tab1}). The refined Fe-site occupancy is slightly larger than 100\%. This might be caused by the presence of small amount of vacancies on Se sites. If one rescales the Rb stoichiometry assuming exactly two Fe per formula  one gets $\rm {Rb_{0.55(5)}}$. The phase separation on two phases was reported in NMR study~\cite{2012arXiv1203.1834T} with composition of the minority vacancy disordered phase $\rm {Rb_{0.3}(Fe Se)_2}$ deduced from intensity measurements. The crystal~\cite{2012arXiv1203.1834T} had similar superconducting transition temperature $T_c=32$~K and the strong difference in Rb-stoichiometry looks rather surprising if one assumes that the minority phase is responsible for SC.

\begin{table}[tb!]
\caption{Crystal structure parameters. The positions in the average structure model $I4/mmm$ (no. 139) are Fe in $(0,{1\over2},{1\over4})$ (4d), Se in $(0,{0},{z})$ (4e) and Rb in $(0,{0},{0})$ (2a). In the  vacancy ordered $I4/m$ (no. 87), whose unit cell is generated by the transformation given in the text the atoms are split in the following way: Rb in  Rb1 $(x=0.4,y={0.8},0)$ (8h) and Rb2 $(0,0,0)$ (2a), Se in Se1  $(x=0.4,y={0.8},z=-z_{Se})$ (16i) and Se2  $({1\over2},{1\over2},z=-z_{Se}+{1\over2})$ (4e), Fe in Fe1 $({1\over2},0,{1\over4})$ (4d) and Fe2 $(x={0.3},y={0.6},z=0.25)$ (16i) , where $z_{\rm Se}$ is z-component in $I4/mmm$. The values after $=$ indicate the ideal undistorted coordinates. The parameters of the main phase are given for T=2~K and above the order-disorder transition at T=560~K with $I4/m$ and $I4/mmm$ space groups, respectively. The structure of the secondary phase is given at 2~K with $I4/mmm$ symmetry.
The occupancies o-Rb1, o-Rb2, o-Fe1 and o-Fe2 are standard site occupancies (SOF) of the respective sites. The stoichiometries s-Rb1, s-Rb2, s-Fe1 and s-Fe2 correspond to the occupancies normalized according with the site multiplicity to be in units of the formula \feserb. Atomic displacement parameters $B$ ($\rm\AA^2$) were constrained to be the same for the same atom types. $\chi^2$ and $R_{wp}$ are the global chi-square and weighted R-factor for the whole pattern.  The magnetic moment M-Fe is in $\rm\mu_B$ and calculated per Fe2 site (i.e. assuming o-Fe2=1). }
 \label{tab1}

\begin{center}
\begin{tabular}{llll}

          &  1 $I4/m$   & 2  $I4/mmm$   &  1  $I4/mmm$  \\
          &    2~K       &   2~K         &   560~K       \\ \hline
 a, \AA        & 8.75395(14)  &  3.8322(2)   &  3.97940(8)     \\    
 c, \AA        & 14.47890(25) & 14.629(1)    &  14.6008(5)     \\

x-Rb1     & 0.3880(16)  &    0         &    0           \\
y-Rb1     & 0.787(2)    &    0         &    0           \\

x-Fe2     & 0.2994(7)   &    0         &    0           \\
y-Fe2     & 0.5933(7)   &    0.5       &    0.5         \\
z-Fe2     & 0.2458(4)   &    0.25      &    0.25        \\

x-Se1     & 0.3905(9)    &    0         &    0          \\
y-Se1     & 0.7953(11)   &    0         &    0          \\
z-Se1     & 0.6465(3)    &   0.3535     &  0.3517(3)       \\
z-Se2     & 0.1507(10)   &   -          &   -           \\

 o-Rb1    & 0.796(16)   &  0.61(5)      & 0.8400(26)          \\
 o-Rb2    & 0.792(48)   &  -            &   -             \\
% sf-Rb1    & 0.398(8)    &  0.61(5)      & 0.4200(13)          \\
% sf-Rb2    & 0.099(6)    &  -            &  -               \\
 s-Rb1    & 0.637(13)    &  0.61(5)      & 0.672(2)          \\
 s-Rb2    & 0.16(1)      &  -            &   -              \\
 s-Rb$_\Sigma$     & 0.79(1)      &  0.61(5)      & 0.8400(26)       \\

% sfullprof-Fe1    & 0.041(4)    &     -       &   -            \\
% sf-Fe2    & 0.894(9)    &  2.2(1)      &  0.773(7)        \\

 o-Fe1    & 0.164(16)    &     -       &        -       \\
 o-Fe2    & 0.894(9)    &  1.10(5)      &  0.773(7)        \\
 s-Fe1    & 0.066(6)    &     -       &       -        \\
 s-Fe2    & 1.43(1)    &  2.2(1)      &  0.773(7)        \\
 s-Fe$_\Sigma$     & 1.496(12)  &  2.2(1)      &  1.546(14)        \\

B-Fe      & -0.02(5)    & -0.02(5)     & 2.84(13)          \\
B-Se      & 0.82(6)     & 0.82(6)      & 4.2(3)           \\
B-Rb      & 1.58(16)    & 1.58(16)     & 2.26(11)       \\
M-Fe2     & 2.72(46)    & 0            & 0         \\

$R_{wp}$,\%& \multicolumn{2}{c}{6.6}& 6.27          \\
$\chi^2$&    \multicolumn{2}{c}{7.1}&    2.63             \\ 
%$R_{B}$,\%&    4.7       &  6.2             &    12             \\ 
\hline
\end{tabular}
\end{center}
\end{table}

\subsection{Temperature dependence of phase separation}

\def\dd{\mm{(\delta d_{st}/d)}}

\begin{figure}[tb!]
  \begin{center}
    \includegraphics[width=8cm]{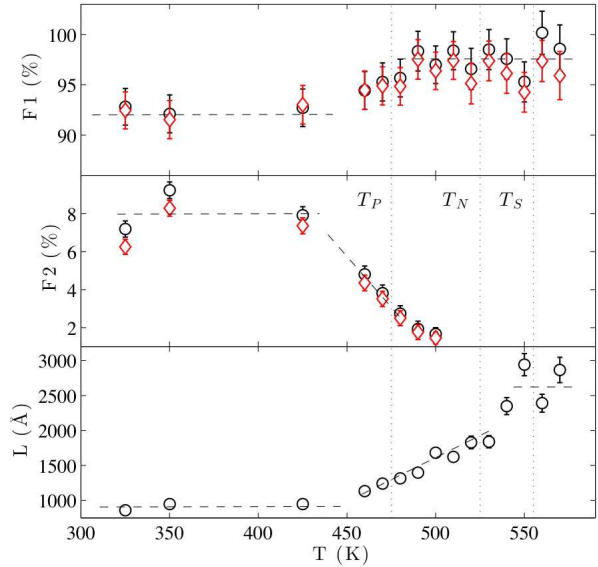} %/Users/pomjakushin/work/Fits/kat/fese/2011_fesex/tdepfigs/fig_fractions_size.m
  \end{center}

 \caption{
Phase fractions of main phase F1 and the the second minority MCP phase F2, calculated from the row refined phase scale factors as explained in the text. The phase fractions have been normalized to correspond to the mass phase fractions at T=325~K. At other temperatures the sum $F1+F2$ might not preserve 100\%. The circles and the rhombs correspond to two slightly different models of constrained and unconstrained refinements of Rb occupancies, respectively. See text for details. $L$ are the spatial dimensions of the phase regions, calculated from the Lorentzian Bragg peak broadening.}
  \label{scaleL}
\end{figure}

To follow the temperature dependence of the phase fractions we calculate the phase fractions from the scale factors $S$. The overall phase scale factor $S$ is a coefficient in front of the calculated $|F(\mathbf{H})|^2$, where $F(\mathbf{H})$ are the unit cell structure factors. $S$ is proportional to $N/v_0$, where N is the number of unit cells (or amount of the phase material), $v_0$ is the unit cell volume \cite{LoveseyI}. Figure \ref{scaleL} shows the refined overall scale factors $S1$ and $S2$ multiplied by the respective unit cell volumes for both phases. This product is proportional to the amount of the phase in the neutron beam. For convenience, the $S(T)v_0(T)$ values were normalized to be in units of mass fraction for the temperature $T=325$~K, assuming that the two phases total 100\%.  For other temperatures no normalization was done, so the sum of the phase fractions F1 and F2 (Fig.~\ref{scaleL}) might not preserve 100\%. We did not perform the normalization intentionally to see where the minority phase transforms on heating. One can see from Fig.~\ref{scaleL} that the minority phase fraction starts to decrease at 470~K and reaches less than 2\% level above 490~K. At the same time the amount of the main phase is increased, implying that the minority phase transforms to the main vacancy ordered phase. On cooling back to the room temperature the second MCP phase appears again apparently due to the phase separation at $T_P=475$~K. The first DSC peak at $T_P$ (Fig.~\ref{dsc}) can be identified as the temperature of the phase separation (or phase merging). Figure \ref{scaleL} has two sets of experimental points. We had noticed that the refined scale factors are slightly model dependent and give the values for both constrained and unconstrained refinements of Rb occupancies, as we mentioned above. An independent sign of the structure transition at $T_P$ comes also from the diffraction peak broadening that can have two contributions. The microstrains or the static fluctuations of crystal lattice constants \dd\ result in $2\dd \tan(\theta)$ peak width dependence.  The apparent sizes  $L$ of the coherently scattered domains of the phases produce a Lorentzian peak broadening with $\sigma_L/\cos(\theta)$ dependence of scattering angle $2\theta$, where $L=\lambda/\sigma_L$. The refined microstrain values were found to be temperature independent with $\dd=0.3$\%. The apparent domain sizes show well pronounced dependence of temperature above the phase separation transition $T_P$. One can see (Fig.~\ref{scaleL}) that the sizes $L$ start to increase concomitantly with the decrease in the fraction of the second phase $F_2$. It is obvious that the refined peak widths are dominated by the main phase and the narrowing of the peaks cannot be a side effect of the disappearance of the MCP contribution from the ND pattern. The sizes $L$ are becoming larger on heating above $T_P$ apparently due to the merging of the regions of the main phase that were separated by the second minority MCP phase at low temperatures.

\subsection{Temperature dependence of the structures}

\begin{figure}[tb!]
  \begin{center}
    \includegraphics[width=8cm]{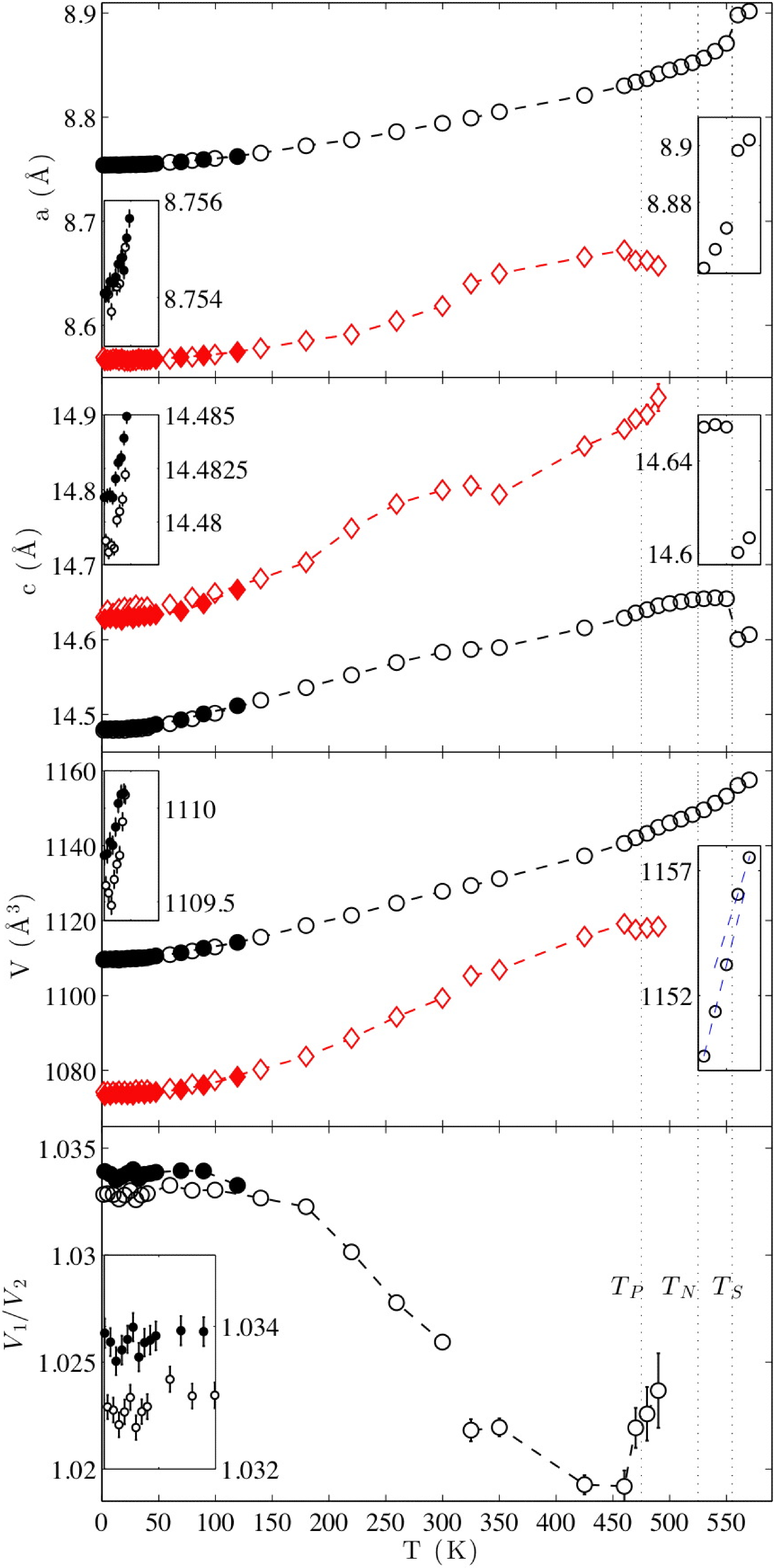}% figure8_metrics_all.\extgra} %/Users/pomjakushin/work/Fits/kat/fese/2011_fesex/tdepfigs/fig_abc_low.m
  \end{center}

 \caption{
Lattice constants and unit cell volume of the main phase (black circles) and the second MCP (red rhombs). Lattice constants $a$ and the volume $V_2$ of the second phase were multiplied by the factors $\sqrt{5}$ and 5, respectively. The data below T=325~K were collected separately from the data at $T\ge325~K$ with the use of samples \rbl\ and \rbh, respectively. All the data were taken on heating. The filled symbols correspond to the second dataset taken just after finishing the first one at room temperature. The bottom plot shows the ratio of the unit cell volumes as a function of temperature. The insets show a zoomed plots for the main phase. The insets have the same $x$-axis as the main plot but different $y$-axes as indicated by tick labels.}
  \label{tabc}
\end{figure}

The temperature dependence of the lattice constants is shown in Fig.~\ref{tabc} for both phases. One can see an abrupt shrinking of the lattice constant $c$ and an expansion in $ab$ plane of the main phase at the structural order-disorder transition at $T_S$ similar as reported in Ref.~\cite{pom2011fesex,Svitlyk2011}. The second phase is compressed in the $ab$-plane and expanded along $c$-axis.  The main phase has 3.3\% bigger unit cell volume at $T=2$~K. The phases have quite different temperature dependences of the lattice constants and unit cell volumes which are also shown in the Fig.~\ref{tabc}. The volume of the second phase increases significantly faster with the temperature increase above $T\simeq150$~K. The lattice constants of MCP phase at the temperatures above $T_P$ might be not accurate due to its very small fraction at these temperatures.

\begin{figure}[tb!]
  \begin{center}
    \includegraphics[width=8cm]{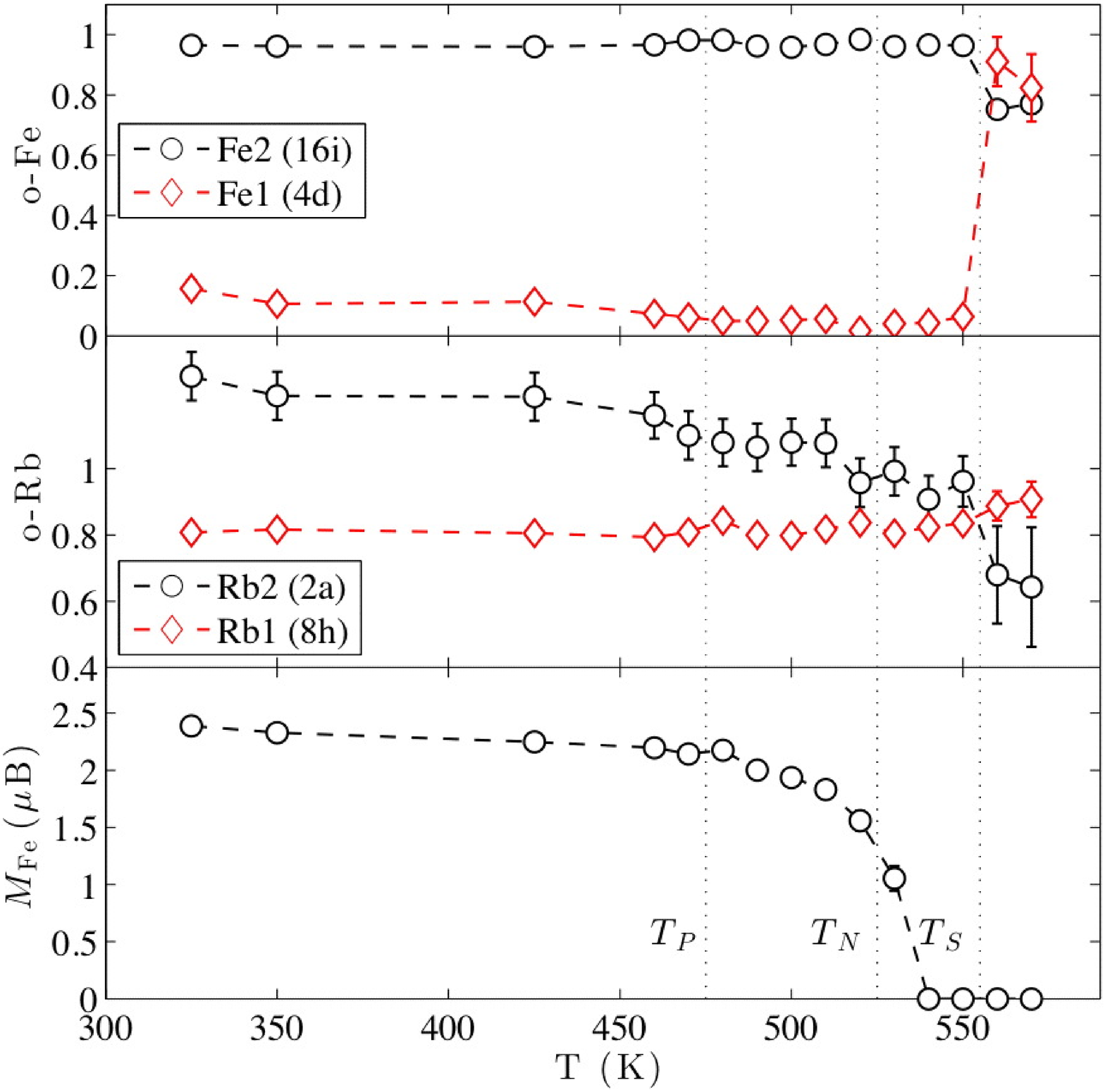} %{figure7_occmom_c.\extgra}{figure7_occ_cs21fix_mom_c.\extgra} %/Users/pomjakushin/work/Fits/kat/fese/2011_fesex/tdepfigs/fig_occmom_c.m
  \end{center}

 \caption{Temperature dependence of site occupancies o-Fe and o-Rb and magnetic moment $M_{\rm Fe}$ of Fe2 site of the main vacancy ordered phase measured above room temperature with the sample \rbh. The moment values are per Fe2 site (i.e. assuming o-Fe2=1). The refinements were performed in the $I4/m$ model for all temperatures.}
  \label{thigh}
\end{figure}

The Fe and Rb occupancies and magnetic moment on Fe at the temperatures above room temperature are shown in Fig.~\ref{thigh}. As we noted above the occupancy of the Rb2 (2a) above unity is probably related to some correlation effects and lower statistics of the experimental data of the temperature scan. However we prefer to use this unconstrained model to see the trends of the changes for all occupancies. Above $T_S$ the occupancies of two sites of Fe are strongly correlated if refined in the $I4/m$ structure and show large errorbars. Nevertheless this type of fit very well shows a jump-like leveling of the Fe-occupancies due to structure order-disorder transition at $T_S$. The magnetic moment gets nonzero values ($<1\mu_B$) if refined above $T_N$ due to correlation with the crystal structure parameters. The refinements were made in two phase model below 500~K and in one phase model (only main $I4/m$ phase) above 500~K.  

\subsection{Irreversibility effects below room temperature }

The low temperature scan with \rbl-sample was first performed from 2K up to room temperature and then the second time after cooling to 2~K again up to 120~K. The closed symbols in Fig.~\ref{tabc} show the second scan datapoints. The lattice constant $a$ kept the same value in both scans, whereas the lattice constants $c$ changed significantly in the second temperature scan in both phases. The lattice constant $c$ of the main phase has increased by $0.015(2)$\%, whereas the constant $c$ of the second MCP phase has oppositely decreased by $0.09(1)$\%, resulting in an overall change of the relative unit cell volumes by $0.10(2)$\% which is well seen in Fig.~\ref{tabc}. This type of behavior is difficult to rationalize in a single phase system. We believe that the irreversibility is reflecting a metastable character of both phases in the phase separated state. When the second phase appears from the main phase on cooling below $T_P$ it has 2\% smaller unit cell volume (Fig.~\ref{tabc}). Since the second phase is created along the whole single crystal volume without the destruction (pulverization) of the crystal it is possible that there are some unrelaxed strains/pressure along the phase boundaries and the resulting state is not a thermodynamically ground state. Further on cooling, the forces acting on the phase boundaries are changed due to the change in the relative phase volumes $V_1/V_2$. As a result of such cycling the strains on the phase boundaries can be relaxed and the phases get closer to the equilibrium state with the larger unit volume of the main phase and the smaller MCP phase volume. 

Hypothetically, one could argue that the transition at $T_P$ is not a phase separation transition, but a  second crystal structure transition to a new structure which we simply could not identify. This new structure below $T_P$ should have yet bigger superstructure unit cell to accomodate all the Bragg reflections observed in the experiment. However, the presence of the irreversibility favors the non-single phase state of the crystal.

\subsection{Change of composition after annealing at $T_P$}
\label{sec_an}
\begin{figure}[tb!]
  \begin{center}
    \includegraphics[width=9.3cm]{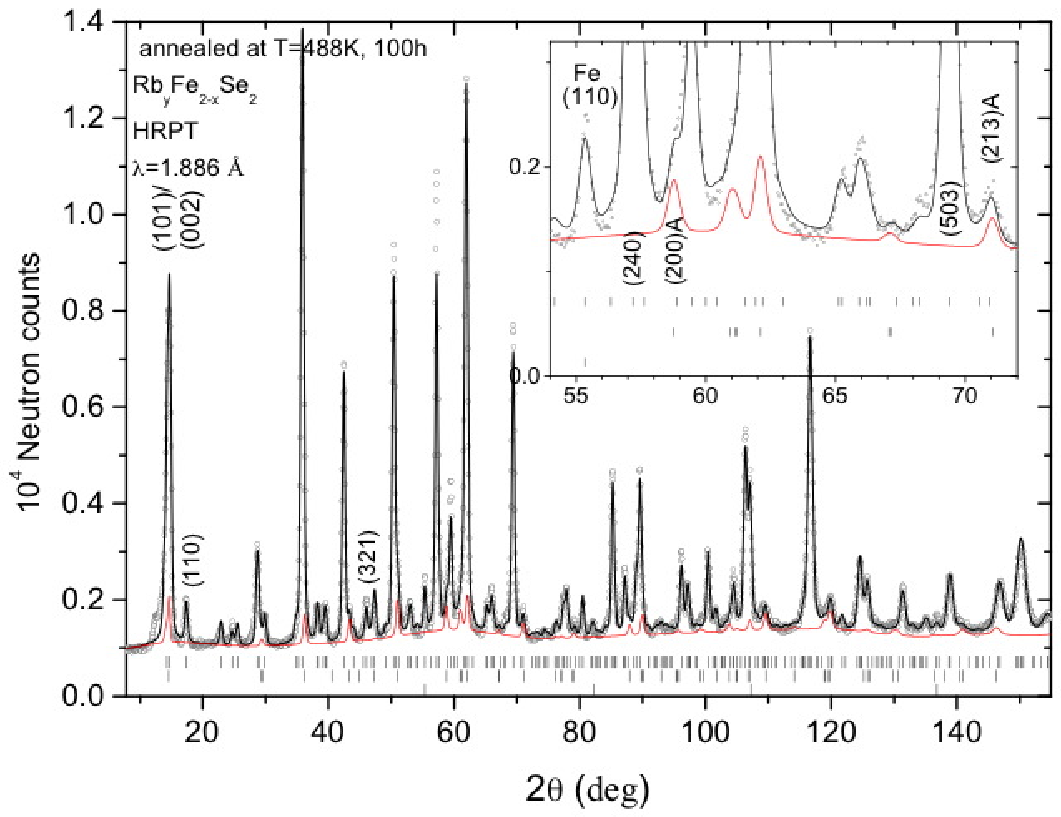} % fesex_riet
  \end{center}

\caption{Room temperature neutron diffraction pattern of \feserb\ \rbl\ after annealing at T=488~K during 100~h. Contribution of the minority in-plane compressed phase is shown by red solid curve. All notations in the figure are the same as in Fig. \ref{rietv-ND}.}
  \label{rietv-NDa}
\end{figure}

To further check the two phase model we have performed annealing of both samples at the temperature $T=488$~K in the vicinity of the phase separation temperature $T_P$ as described in section \ref{exp}. After that the room temperature neutron powder diffraction patterns were collected. Figure \ref{rietv-NDa} shows the diffraction pattern of the annealed \rbl\ sample. One can see that the amount of the second MCP phase is drastically decreased in comparison with the original sample K72 (Fig.~\ref{rietv-ND}). At the same time the amount of the Fe-impurity phase has increased as well seen from the $(110)$ peak intensity at $2\theta=55.43^o$ (the intensity at this $2\theta$-position  has practically zero contribution from the main phase). In addition to the decrease in the amount of the MCP, its structure becomes further more in-plane compressed ($a=5.83$~\AA) after annealing as also well seen from the shift of the peak $(200)_A$ in Fig.~\ref{rietv-NDa} in comparison with Fig.~\ref{rietv-ND}.  The refined minority phases content has changed from 10.9\% MCP, 0.3\% Fe (mass fraction) for the original K72 to 5.8\% MCP, 1.4\% Fe for the annealed K72, and from 8\% MCP, $<0.3$\% Fe for the original K43 to 4.7\% MCP, 1.1\% Fe for the annealed K43. This observation provides additional evidences in favor of metastable character of the minority phase. The appearance of elemental Fe is in accordance with the larger stoichiometry of the iron in the minority phase Fe2.2 in comparison with the main AFMV phase Fe1.6.    

\section{Summary}

The crystal and magnetic structures of the superconducting \feserb\ have been studied by means of single crystal synchrotron x-ray and neutron powder diffraction (NPD) and differential calorimetry (DSC) in the temperature range from 2-570~K. The ground state of the crystal is an intrinsically phase separated state with two distinct by symmetry phases. The main phase possesses the iron vacancy ordered \xfive\ superstructure ($I4/m$ space group) with AFM ordered Fe spins. The minority phase does not have \xfive-type of ordering and has smaller in plane lattice constant $a$ and larger tetragonal $c$-axis. The NPD data can be very well described assuming the parent average vacancy disordered structure ($I4/mmm$ space group) of the minority phase with the refined stoichiometry  $\rm {Rb_{0.60(5)}(Fe_{1.10(5))}Se})_2$. The minority phase amounts to 8-10\% mass fraction. The unit cell volume of the minority phase is 3.2\% smaller than the one of the main phase at $T=2$~K and has different temperature dependence. We note that the true crystal symmetry of the minority phase might be lower than tetragonal but due to its small amount and powder averaging the average $I4/mmm$ approximation works very well.

The minority phase fraction, calculated from structure refinements of the NPD data, decreases at phase separation temperature $T_P=475$~K and reaches less than 2\% level above 490~K. At the same time the amount of the main phase is increased, providing the direct evidence that the minority phase transforms to the main vacancy ordered phase. The spatial dimensions of the phase domains strongly increase above $T_P$ due to the merging of the regions of the main phase that were separated by the second minority MCP phase at low temperatures. The phase transition of the pure main phase to the vacancy disordered structure occurs at higher temperatures $T>525$~K. 

The phase separation, the antiferromagnetic and the order-disorder transition  temperatures observed in NPD experiment have the respective peaks in DSC calorimetry temperature scans.

\section*{A{\lowercase{cknowledgements}}}

The authors acknowledge the allocation of the beam time at Swiss-Norwegian beam line (BM1A) of the European Synchrotron Radiation Facility (ESRF, Grenoble, France). The authors thank the NCCR MaNEP project and Sciex-NMS$\rm ^{ch}$ (Project Code 10.048) for the support of this study. The work was partially performed at the neutron spallation source SINQ.

\section*{References}

\bibliography{../../../refs/pomjakus_webofknowledge,../../../refs/refs_general,../../../refs/refs_fesex,../../../refs/refs_manganites,../../../publication_list/publist2007,../../../publication_list/publist2006,../../../publication_list/publist2005,../../../publication_list/publist}

\end{document}